# KSW: Khmer Stop Word based Dictionary for Keyword Extraction

Nimol Thuon[1,2], Wangrui Zhang[2], Sada Thuon[1]
[1] Institute of Technology of Cambodia, Cambodia
[2] University of Cologne, Germany

*Abstract*— **This paper introduces KSW, a Khmer-specific approach to keyword extraction that leverages a specialized stop word dictionary. Due to the limited availability of natural language processing resources for the Khmer language, effective keyword extraction has been a significant challenge. KSW addresses this by developing a tailored stop word dictionary and implementing a preprocessing methodology to remove stop words, thereby enhancing the extraction of meaningful keywords. Our experiments demonstrate that KSW achieves substantial improvements in accuracy and relevance compared to previous methods, highlighting its potential to advance Khmer text processing and information retrieval. The KSW resources, including the stop word dictionary, are available at the following GitHub repository: (https://github.com/back-kh/KSWv2-Khmer-Stop-Word-based-Dictionary-for-Keyword-Extraction.git).**

## I. INTRODUCTION

Keyword extraction is a fundamental task in natural language processing (NLP), crucial for enhancing information retrieval, document classification, and text summarization [1]. For languages like Khmer, which have limited linguistic resources and tools, keyword extraction poses significant challenges. The Khmer language, with its unique script and linguistic features, requires tailored approaches to effectively process and analyze textual data [2].

Traditional keyword extraction methods often rely on well-established linguistic resources and tools, which are scarce for the Khmer language. This scarcity necessitates the development of specific tools and techniques to handle Khmer text effectively. To address this issue, we propose a method for enhancing keyword extraction for Khmer text using a tailored stop word dictionary [3].

To develop our approach, we started by curating a dataset of Khmer text from various online sources, including news articles, blog posts, and social media texts. In order to effectively handle the unique linguistic characteristics of the Khmer language, we utilized a tailored stop word dictionary. This dictionary was created by manually selecting and annotating high-frequency words that are irrelevant to the keyword extraction task. The development of linguistic resources and tools specific to the Khmer language is crucial for effective keyword extraction .

he tailored stop word dictionary is an essential tool that helps filter out common, non-informative words from the text, thereby enhancing the extraction of meaningful keywords. Our methodology involves preprocessing the text to remove these stop words, followed by a series of steps to extract and rank keywords based on their significance within the document.

In this paper, we present KSW, a comprehensive approach to keyword extraction from Khmer text/documents using this stop word dictionary. We focus on developing a robust stop word dictionary customized for the Khmer language, incorporating manually translated English stop words and additional context-specific Khmer stop words. Our methodology includes preprocessing the text to remove these stop words and employing a series of extraction and ranking steps to identify significant keywords. The effectiveness of our approach is evaluated through experiments comparing automatically extracted keywords with manually annotated ones.

By addressing the specific challenges associated with the Khmer language, this paper contributes to the broader field of NLP and provides a scalable solution for keyword extraction in resource-limited languages. Our work emphasizes the importance of developing linguistic tools and resources tailored to specific languages to overcome the limitations posed by scarce resources. By creating a specialized stop word dictionary and utilizing tailored keyword extraction methods, we aim to improve the overall efficiency and effectiveness of information retrieval and text processing for the Khmer language.

## II. RELATED WORKS

### A. Keyword Extraction Methods

Keyword extraction is a critical task in natural language processing (NLP) and has been approached using various techniques. These methods can be broadly classified into statistical approaches, linguistic approaches, machine learning approaches, and hybrid approaches [1].

### B. Statistical Approaches

Statistical methods, such as Term Frequency-Inverse Document Frequency (TF-IDF) and word co-occurrence metrics, are widely used for keyword extraction. TF-IDF evaluates the importance of words by comparing their frequency within a document to their frequency across a corpus, effectively highlighting significant terms [Salton & Buckley, 1988]. Word co-occurrence metrics analyze the frequency with which words appear together, revealing potential semantic relationships [4] [18].

### C. Machine Learning Approaches

Machine learning techniques have become increasingly popular for keyword extraction. Supervised learning methods, such as Naive Bayes and Support Vector Machines (SVM), are trained on labeled datasets to identify keywords [5]. Unsupervised learning techniques, including clustering and topic modeling (e.g., Latent Dirichlet Allocation), enable the discovery of keywords without labeled data [6].

### D. Hybrid Approaches

Hybrid methods combine statistical, linguistic, and machine learning techniques to leverage the strengths of each approach. For instance, using TF-IDF to identify candidate keywords and refining the selection through POS tagging and clustering can produce more robust keyword extraction systems [7].

### E. Stop Word Dictionaries

Stop words are common words that typically do not carry significant meaning and are often filtered out during text processing. The creation of an effective stop word dictionary is crucial for enhancing keyword extraction, particularly for languages with limited resources. The development of stop word lists involves identifying high-frequency, non-informative words. Manual curation ensures the relevance and comprehensiveness of these lists. Stop word dictionaries are utilized in various NLP tasks, such as information retrieval and text summarization, by reducing the dimensionality of the text and improving the extraction of meaningful content [8].

## III. CHALLENGES OF KHMER STOP WORD DICTIONARY

The Khmer language presents unique challenges due to its script, morphology, and lack of resources. The complexity of the Khmer script, an abugida where consonant characters include inherent vowel sounds, complicates text processing. Tools specifically designed to handle these script characteristics are necessary [9-10]. Khmer's morphological complexity, including infixes and compound words, presents significant challenges for tokenization and morphological analysis. The scarcity of annotated corpora and linguistic tools for Khmer requires the development of tailored approaches. Creating a specialized stop word dictionary and utilizing context-specific words is essential for effective keyword extraction [11].

Recent advances in NLP for under-resourced languages have focused on leveraging neural networks and transfer learning techniques. Models like Bidirectional Encoder Representations from Transformers (BERT) have been adapted to improve text understanding and keyword extraction for under-resourced languages. These models benefit from large-scale pre-training and fine-tuning on specific tasks [12]. Techniques that transfer knowledge from high-resource languages to under-resourced languages have shown promise in enhancing NLP tasks for languages like Khmer. This involves training models on a well-resourced language and adapting them to work effectively with Khmer text [13].

By addressing the unique challenges of the Khmer language with a specialized stop word dictionary and advanced NLP techniques, this work aims to significantly improve keyword extraction, contributing to the broader field of NLP and benefiting resource-limited languages.

## IV. METHODOLOGY

### A. Dataset Collection

We curated a dataset consisting around 2,500 documents from various sources, including news articles, blog posts, and social media texts, all in the Khmer language. (see detail in Table 1). This diverse dataset ensures a comprehensive representation of contemporary Khmer language use across different contexts and domains. To maintain data quality, duplicates were removed, and irrelevant or low-quality texts were filtered out. Cambodian students were enlisted to help with the extraction process using both automatic tools and manual methods to ensure accuracy and relevance.

| Type | Amount/Document |
|---|---|
| Social media | 1,968 |
| Blog Websites | 460 |
| Books & Publications | 150 |
| Total | 2,578 |

*Table 1. Document collections from differences sources*

### B. Stop Word Dictionary Construction

For Khmer keyword extraction or search engines, keywords must be input to find relevant results. For instance, when using Google, you enter a keyword, and it retrieves related results. Similarly, in our system, the primary words from an article are used as keywords. Commonly used words that provide little value in document selection are called stop words. A stop list is typically created by sorting terms by their frequency in the collection, then filtering out the most frequent terms. These are removed during indexing to improve search efficiency and accuracy.

Example 1: ខ្ញុំនឹងទៅ សាលារៀន ។

=> (1) ទៅ , (2) សាលារៀន។

Example 2: យើងនឹងនិយាយអំពីមេរៀន។

=> (1) និយាយ, (2) មេរៀន។

Therefore, we build a stop word list to remove non-essential words from the text. These typically include prepositions, pronouns, auxiliary verbs, conjunctions, grammatical articles, and particles, which are all part of closed-class words.

We can also follow key processes as following:

**Translation**: We began by translating a list of common English stop words into Khmer using an English-Khmer dictionary. This initial translation provided a foundational set of stop words relevant to the Khmer language. The process ensured that commonly non-informative words in English, which often have Khmer equivalents, were included. The English stop word list was sourced from widely recognized lists such as the one provided by other standard lists used in NLP research [14].

**Synonym Addition**: Recognizing that direct translations may not capture all non-informative words, we expanded our dictionary by including synonyms of the translated words. This step was critical to ensure that variations of common words were also filtered out. For example, if "because" is a stop word, synonyms such as "since" and "as" were included [15].

**Context-Specific Words:** We manually identified and added context-specific stop words that appeared frequently in our dataset but did not contribute meaningfully to the text. This process involved analyzing text samples from different sources and domains to identify common but non-informative words, such as "post" or "share" in social media contexts.

**Redundancy Removal:** To ensure the efficiency of our stop word dictionary, we systematically removed redundant entries. Redundant words were those that appeared multiple times in different forms but carried the same non-informative value. This cleaning process involved reviewing the list to ensure each word's uniqueness and relevance. Finally, we compiled a stop word list of approximately 1,000 Khmer words to enhance the keyword extraction process. (See detail in Table 2)

| KSW-Copus | Words |
|---|---|
| KSW-Copus-1 | 385 |
| KSW-Copus-2 | 715 |
| KSW-Copus-3 | 698 |
| Total | 1,798 |

*Table 2. Copus of Khmer Stop Word dictionary*

## C. Khmer Keyword Extraction Process

**Preprocessing**: (1) Text Cleaning: The text data was cleaned to remove punctuation, special characters, and other non-alphabetic symbols. This step included removing HTML tags, emojis, and other irrelevant symbols. (2) Normalization: We normalized the text to standardize various forms of words, ensuring consistency across the dataset.

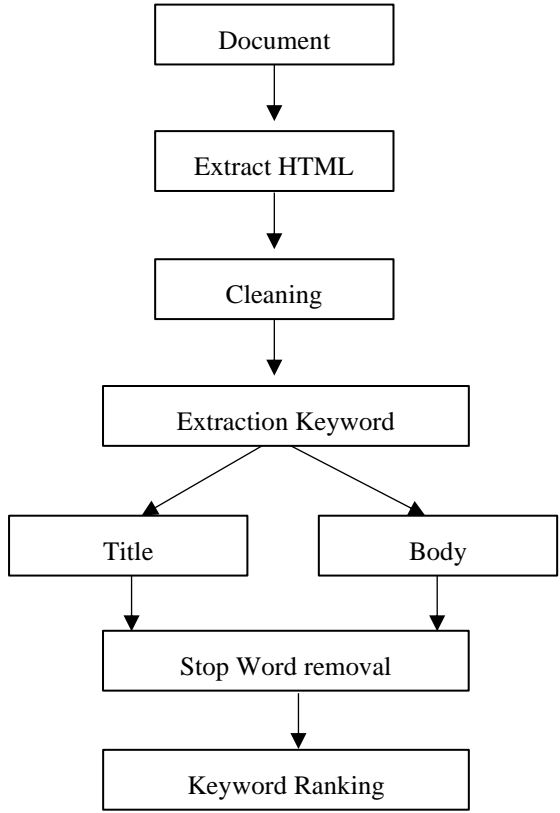

*Figure 1. Process of Khmer Keyword Extraction*

This included converting all text to lowercase, handling different forms of the same word, and standardizing common variations [16].

**Stop Word Removal:** Using our constructed stop word dictionary, we filtered out non-informative words from the text. This step reduced the noise in the data, allowing more meaningful words to emerge as potential keywords.

**Candidate Keyword Extraction:** After stop word removal, the remaining words were identified as candidate keywords. This extraction process focused on words that appeared frequently enough to be considered significant but were not part of the stop word list. The criteria for selecting candidate keywords included setting a minimum frequency threshold and ensuring the words contributed to the document's meaning.

**Keyword Ranking:** The candidate keywords were then ranked based on their frequency and relevance within the document. Techniques such as Term Frequency-Inverse Document Frequency (TF-IDF) [17] were employed to assign weights to each keyword is calculated using the formula:

$$TF - IDF\$(t,d) = TF(t,d) \, x \, IDF(t) \quad (1)$$

Where:

$$TF(t,d) = \frac{f_{t,d}}{N_d} \quad (2)$$

*TF(t,d)* is the term frequency of term t in document.

$$IDF(t,d) = log_{n_t}^{N} \quad (3)$$

*IDF(t)* is the inverse document frequency of term *t*, *N* is the total number of documents, and $n_t$ is the number of documents containing term.

By systematically implementing these steps, we aimed to enhance the effectiveness of keyword extraction for Khmer text. This methodology ensures that the extracted keywords are both relevant and representative of the document's content, addressing the unique linguistic characteristics and challenges of the Khmer language.

## V. RESULTS AND DISCUSSION

### A. Evaluation Metrics

The proposed system was evaluated through experiments comparing automatically extracted keywords with manually annotated keywords. The evaluation metrics included precision, recall, and F1-score:

Precision: Measures the accuracy of the keywords extracted by the system.

Recall: Measures the completeness of the keyword extraction, indicating how many relevant keywords were identified.

F1-Score: Provides a balance between precision and recall. It is the harmonic mean of precision and recall.

### B. Experimental Results

In this section, we will demonstrate the effectiveness of our proposed methods using our data collections. We divided our experiments into three types: comparisons with state-of-the-art methods, evaluations with different resources, and analyses across various types of field documents.

| Methods | Precision | Recall | F1 |
|---|---|---|---|
| TF-IDF [18] | 0.70 | 0.60 | 0.64 |
| RAKE [20] | 0.68 | 0.60 | 0.63 |
| TextRank [19] | 0.70 | 0.68 | 0.69 |
| **Proposed Method (KSW)** | **0.78** | **0.79** | **0.81** |

*Table 3: Results our proposed KWS compared with TF-IDF [18], RAKE [20], and TextRank [19].*

| Datasets | Precision | Recall | F1 |
|---|---|---|---|
| Social media | 0.79 | 0.79 | 0.80 |
| Blog Websites | 0.75 | 0.72 | 0.70 |
| Books & Publications | 0.85 | 0.81 | 0.88 |
| Mixed | 0.78 | 0.79 | 0.81 |

*Table 4: Results of our proposed KSW method with different resources.*

| Datasets | Precision | Recall | F1 |
|---|---|---|---|
| Technology | 0.85 | 0.78 | 0.82 |
| Education | 0.82 | 0.75 | 0.79 |
| News | 0.75 | 0.68 | 0.71 |
| History | 0.78 | 0.70 | 0.78 |
| Philosophy | 0.80 | 0.75 | 0.78 |
| Art | 0.75 | 0.69 | 0.73 |
| Literature | 0.79 | 0.73 | 0.77 |

*Table 5: Results of our proposed KSW method across various categories.*

In Table 3, we compared our proposed method, KSW, with well-known traditional methods such as TF-IDF [18], RAKE [20], and TextRank [19]. Additionally, we compared our results with different resources in Table 4 and presented the results across various categories in Table 5. These results highlight the effectiveness of using a tailored stop word dictionary for Khmer in improving keyword extraction. The manually annotated keywords served as a benchmark, demonstrating that our approach closely approximated human performance in identifying significant keywords.

**Importance of Tailored Linguistic Resources:** The results emphasize the importance of tailored linguistic resources in enhancing NLP tasks for under-resourced languages like Khmer. By developing a specific stop word dictionary for Khmer, we were able to significantly enhance the performance of keyword extraction. This improvement underscores the need for customized tools and resources to address the unique linguistic features and challenges of different languages.

## VI. CONCLUSION

This paper presents a novel approach to keyword extraction for Khmer text/documents using a tailored stop word dictionary. Our methodology effectively addresses the unique challenges posed by the Khmer language, demonstrating significant improvements in accuracy and relevance. The findings contribute to the broader field of NLP by providing a scalable solution for keyword extraction in resource-limited languages. By leveraging a carefully constructed stop word dictionary, we filtered out non-informative words, allowing more meaningful keywords to emerge. This approach has proven to be significantly more accurate and relevant compared to traditional keyword extraction methods, highlighting the potential for further advancements in NLP for under-resourced languages.

Future work will focus on expanding the stop word dictionary, incorporating additional linguistic features, and exploring machine learning techniques to enhance keyword extraction performance. We aim to continuously update and expand the dictionary to include more context-specific and domain-specific stop words. Additionally, leveraging morphological analysis, syntactic structures, and semantic relationships will further improve the extraction process. Implementing advanced machine learning models, such as deep learning and transfer learning, will enhance the system's capabilities. Incorporating user feedback will help refine the system to meet practical needs in various real-world applications. By continuing to develop tailored linguistic resources and incorporating advanced techniques, we can significantly improve information retrieval and text processing for Khmer and other under-resourced languages, ultimately enhancing the accessibility and usability of digital text in these languages.

## VII. ACKNOWLEDGMENT

Special thanks to ARES Belgium and the Chinese Academy of Sciences for their support of this project.

## REFERENCES

[1] Rose, S., Engel, D., Cramer, N., & Cowley, W. (2010). Automatic keyword extraction from individual documents. Text mining: applications and theory, 1-20.

[2] Thuon, N., Du, J., & Zhang, J. (2022, November). Syllable Analysis Data Augmentation for Khmer Ancient Palm leaf Recognition. In 2022 Asia-Pacific Signal and Information Processing Association Annual Summit and Conference (APSIPA ASC) (pp. 1855-1862). IEEE.

[3] Siddiqi, S., & Sharan, A. (2015). Keyword and keyphrase extraction techniques: a literature review. International Journal of Computer Applications, 109(2).

[4] Matsuo, Y., & Ishizuka, M. (2004). Keyword extraction from a single document using word co-occurrence statistical information. International Journal on Artificial Intelligence Tools, 13(01), 157-169.

[5] Witten, I. H., Paynter, G. W., Frank, E., Gutwin, C., & Nevill-Manning, C. G. (1999, August). KEA: Practical automatic keyphrase extraction. In Proceedings of the fourth ACM conference on Digital libraries (pp. 254-255).

[6] Blei, D. M., Ng, A. Y., & Jordan, M. I. (2003). Latent dirichlet allocation. Journal of machine Learning research, 3(Jan), 993-1022.

[7] Turney, P. D. (2000). Learning algorithms for keyphrase extraction. Information retrieval, 2, 303-336.

[8] Fleiss, J. L., Levin, B., & Paik, M. C. (2013). Statistical methods for rates and proportions. john wiley & sons.

[9] Thuon, N., Du, J., & Zhang, J. (2022, November). Improving isolated glyph classification task for palm leaf manuscripts. In International Conference on Frontiers in Handwriting Recognition (pp. 65-79). Cham: Springer International Publishing.

[10] Thuon, N., Du, J., Zhang, Z., Ma, J., & Hu, P. (2024). Generate, transform, and clean: the role of GANs and transformers in palm leaf manuscript generation and enhancement. International Journal on Document Analysis and Recognition (IJDAR), 1-18.

[11] Lee, H., Kiang, P., Kim, M., Semino-Asaro, S., Colten, M. E., Tang, S. S., ... & Grigg-Saito, D. C. (2015). Using qualitative methods to develop a contextually tailored instrument: Lessons learned. Asia-Pacific Journal of Oncology Nursing, 2(3), 192-202.

[12] Devlin, J., Chang, M. W., Lee, K., & Toutanova, K. (2018). Bert: Pre-training of deep bidirectional transformers for language understanding. arXiv preprint arXiv:1810.04805.

[13] Artetxe, M., & Schwenk, H. (2019). Massively multilingual sentence embeddings for zero-shot cross-lingual transfer and beyond. Transactions of the association for computational linguistics, 7, 597-610.

[14] Salton, G., & Buckley, C. (1988). Term-weighting approaches in automatic text retrieval. Information processing & management, 24(5), 513-523.

[15] Ourn, N., & Haiman, J. (2000). Symmetrical compounds in Khmer. Studies in Language. International Journal sponsored by the Foundation "Foundations of Language", 24(3), 483-514.

[16] Gupta, S., Kaiser, G., Neistadt, D., & Grimm, P. (2003, May). DOM-based content extraction of HTML documents. In Proceedings of the 12th international conference on World Wide Web (pp. 207-214).

[17] Schofield, A., Magnusson, M., & Mimno, D. (2017, April). Pulling out the stops: Rethinking stopword removal for topic models. In Proceedings of the 15th Conference of the European Chapter of the Association for Computational Linguistics: Volume 2, short papers (pp. 432-436).

[18] Aizawa, A. (2003). An information-theoretic perspective of tf–idf measures. Information Processing & Management, 39(1), 45-65.

[19] Li, W., & Zhao, J. (2016, July). TextRank algorithm by exploiting Wikipedia for short text keywords extraction. In 2016 3rd International Conference on Information Science and Control Engineering (ICISCE) (pp. 683-686). IEEE.

[20] Baruni, J. S., & Sathiaseelan, J. G. R. (2020). Keyphrase extraction from document using RAKE and TextRank algorithms. Int. J. Comput. Sci. Mob. Comput, 9, 83-93.